\documentclass[10pt]{iopart}
\usepackage[T1]{fontenc}
\usepackage{graphicx}
\usepackage{color}
\usepackage{bm}
\usepackage{stackengine}

\definecolor{darkblue}{rgb}{0,0,0.6}
\definecolor{darkred}{rgb}{0.6,0,0}
\usepackage[colorlinks=true,urlcolor=darkblue,citecolor=darkblue, linkcolor=darkred,hyperfootnotes=false]{hyperref}

\graphicspath{ {figures/} }

\newcommand{\dd}{\mathrm{d}}
\newcommand{\ee}{\boldsymbol{e}}
\newcommand{\EE}{\boldsymbol{E}}

\newcommand{\ff}{\boldsymbol{f}}
\newcommand{\jj}{\boldsymbol{j}}
\newcommand{\JJ}{\boldsymbol{J}}
\newcommand{\kk}{\boldsymbol{k}}
\newcommand{\mcO}{\mathcal{O}}
\renewcommand{\ss}{\boldsymbol{s}}
\newcommand{\uu}{\boldsymbol{u}}

\newcommand{\xx}{\boldsymbol{x}}

\newcommand{\nnu}{\boldsymbol{\nu}}
\newcommand{\zzeta}{\boldsymbol{\zeta}}
\newcommand{\XX}{\boldsymbol{X}}

\newcommand{\ind}[1]{_{\mathrm{#1}}}
\newcommand{\ldeb}{\lambda\ind{D}}
\newcommand{\transp}{^\mathrm{T}}
\stackMath
\newcommand{\underset}[2]{\,\stackunder{#2}{\scalebox{0.8}{$#1$}}\,}

\begin{document}

\title{Stationary and transient correlations in driven electrolytes}

\author{Haggai Bonneau$^{1,2}$,
	Vincent Démery$^{2,3}$ and
	Elie Raphaël$^2$}

\address{$^1$ School of Physics and Astronomy, Tel Aviv University, Ramat Aviv 69978, Tel Aviv, Israel}

\address{$^2$ Gulliver, UMR CNRS 7083, ESPCI Paris PSL, 75005 Paris, France}

\address{$^3$ Univ. Lyon, ENS de Lyon, Univ. Claude Bernard,
	CNRS, Laboratoire de Physique, F-69342, Lyon, France}

\ead{haggai.bonneau@mail.huji.ac.il}

\vspace{10pt}

\begin{indented}
	\item[]\today
\end{indented}

\begin{abstract}
Particle-particle correlation functions in ionic systems control many of their macroscopic properties. In this work, 
we use stochastic density functional theory to compute these correlations, and then we analyze their long-range behavior.
In particular, we study the system's response to a rapid change (quench) in the external electric field.
We show that the correlation functions relax diffusively toward the non-equilibrium stationary state and that in a stationary state, they present a universal conical shape. This shape distinguishes this system from systems with short-range interactions, where the correlations have a parabolic shape.
We relate this temporal evolution of the correlations to the algebraic relaxation of the total charge current reported previously.
\end{abstract}


\maketitle

\section{Introduction}

In many-particle systems with pair interactions, the transport properties, such as viscosity or conductivity, are related to the pair correlation in non-equilibrium steady states.
The conductivity of electrolytes, for instance, has been explored with this approach in the early works of Debye and Hückel~\cite{Debye1923} and later Onsager~\cite{Onsager1927}.
There, the correlations between the ions when the electrolyte is submitted to an external electric field give rise to a negative correction to the Nernst-Einstein conductivity~\cite{Onsager1927,Onsager1957Wien}.
Following the same approach, Onsager and Fuoss computed the effect of the interactions between ions on the viscosity of electrolytes~\cite{onsager1932}.
Apart from electrolytes, the viscosity of a suspension of hard spheres has been deduced from the particle correlations~\cite{brady1997microstructure}.
In all these examples, the difficulty lies in calculating the pair correlations; deducing the transport properties from them is straightforward.

Stochastic Density Functional Theory (SDFT), or Dean-Kawasaki equation~\cite{kawasaki1994, Dean1996}, has recently emerged as a systematic tool to compute the correlations in systems with pair interactions under the Debye-Hückel, or Random Phase, approximation~\cite{hansen2013}.
This framework has been used to generalize the Onsager and Kim results to arbitrary spatial dimensions~\cite{Demery2016Conductivity}, compute the viscosity of a soft suspension~\cite{kruger2018stresses} and account for the finite ion size in the viscosity of an electrolyte~\cite{robin2024correlation}.
It has also allowed to compute the thermal Casimir interaction between plates containing Brownian charges out of equilibrium~\cite{Dean2014, Dean2016} or between dielectric slabs confining a driven electrolyte~\cite{Mahdisoltani2021, du2024correlation}.
SDFT has also been coupled with fluctuacting hydrodynamics, providing a stochastic field theory for the ionic densities and the flow, to account for hydrodynamic interactions between the particles~\cite{Peraud2017,Donev2019}. 
This approach has led to advancements such as accounting for finite ion size using modified interaction kernels, yielding better quantitative predictions of the conductivity at high ionic concentrations~\cite{Avni2022Conductivity, Avni2022Conductance, bernard2023analytical}.

Stochastic Density Functional Theory has thus been widely used to compute the pair correlations in electrolytes and deduce macroscopic transport coefficients from them.
The correlations themselves, hovewer, have received far less attention. 
At equilibrium and under the Debye-Hückel approximation, the correlations between ions in an electrolyte take a Yukawa form and decay exponentially. On the contrary it was noted that they were long-ranged out of equilibrium~\cite{Mahdisoltani2021}.
This situation is reminiscent of driven binary mixtures with short-range interactions, with an overdamped or an active dynamics, which also display algebraic correlations with a  universal parabolic shape~\cite{poncet2017, bain2017critical}.
This raises the question of the shape of the correlations in a driven electrolyte, and in particular of the effect of the long-ranged electrostatic interactions.

Moreover, an algebraic relaxation of the current in an electrolyte following a sudden switching on of the external field has been found recently~\cite{bonneau2023temporal}.
Conversely, it has been found that the current relaxes exponentially following the sudden switching off of the external field.
This peculiar behavior extends the question of the shape of the correlations: what is the shape of the correlations in the transient regime separating equilibrium and a non-equilibrium steady state (NESS)?
How does the shape of the correlations explain the algebraic or exponential relaxation of the electric current?

In this work, we use SDFT with hydrodynamic interactions~\cite{Demery2016Conductivity, Donev2019} to investigate the dynamics of ionic correlations in a bulk electrolyte following a sudden change in the external field.
We first analyze the stationary correlations in NESS and show that, at large distances, the correlations adopt a conical shape, which corresponds to an anisotropic Poisson equation. 
We then address the transient dynamics of the correlations as the system evolves from equilibrium to NESS and back, and find that an anisotropic diffusion equation governs their evolution.
This diffusive scaling of the correlations explains the algebraic relaxation of the macroscopic current observed when going from equilibrium to NESS~\cite{bonneau2023temporal}, and its exponential relaxation from NESS to equilibrium.

The outline of this article is as follows.
In section~\ref{sec:model}, we present the model system and derive general expressions for the ionic correlations; the expressions are later reduced to the binary symmetric electrolyte case. 
In section~\ref{sec:stationary_case}, we analyze the large distance behavior of the correlations in the NESS and derive a universal shape for them.
In section~\ref{sec:time dependent}, we expand the analysis to the temporal response of the correlations to a quench in the driving field at large distances. 
In section~\ref{sec:eq for long range}, we reexamine the problem at the level of the density and charge fluctuations to explain the algebraic temporal relaxation of the charge current.

\section{Model}
\label{sec:model}
We model an electrolyte as a system of charged Brownian particles of different species~\cite{Demery2016Conductivity,Mahdisoltani2021,Peraud2017,Avni2022Conductivity}. The particles move in a homogeneous three-dimensional solution and are subjected to a uniform external electric field with a time-dependent amplitude $\EE(t)=E(t)\hat{\ee}_x$, where $\hat{\ee}_x$ is the unit vector along the $x$ axis. 
The particles interact via the electrostatic potential and are advected by the flow in the solution, which is generated by the forces transmitted by the particles on the solvent.
We denote $\bar\rho_\alpha$ the average density of the particles of the species $\alpha$, $\kappa_\alpha$ their mobility, and $qz_\alpha$ their charge, with $q$ being the elementary charge.
We assume that the system is electroneutral, $\sum_\alpha z_\alpha\bar\rho_\alpha=0$.

We describe the evolution of the density field $\rho_\alpha(\xx,t)$  of the species $\alpha$ using Stochastic Density Functional Theory~\cite{Dean1996,Demery2016Conductivity} with hydrodynamic interactions~\cite{Peraud2017,Donev2019}. 
The density fields are defined as $\rho_\alpha(\xx,t)=\sum_i \delta(\XX_i(t)-\xx)$, where $\XX_i(t)$ is the position variable of the particle $i$ in the species $\alpha$, and the dynamics of $\rho_{\alpha}$ is given by 
\begin{eqnarray}
	\dot\rho_\alpha = - \nabla\cdot \boldsymbol{j}_\alpha, \label{eq:continuity}\\
	\jj_\alpha
	= \boldsymbol{u}\rho_\alpha -T\kappa_\alpha \nabla \rho_\alpha
	+ \kappa_\alpha   \rho_\alpha \boldsymbol{f}_\alpha
	+\sqrt{\kappa_\alpha T \rho_\alpha} \boldsymbol{\zeta}_\alpha, \label{eq:particle_current}
\end{eqnarray}
where $\boldsymbol{u}(\xx,t)$ is the velocity field of the solution, $T$ is the temperature (we set the Boltzmann constant to $k\ind{B}=1$) and $ \boldsymbol{f}_\alpha(\xx,t)$ is the force acting on the particles of the species $\alpha$.
The noise term $\boldsymbol{\zeta}(\xx,t)$ is a Gaussian white noise with the correlation
\begin{equation}\label{eq:noise}
	\langle \boldsymbol{\zeta}_\alpha(\xx,t) \boldsymbol{\zeta}_\beta(\xx',t') \rangle
	= 2 \delta_{\alpha\beta} \delta(\xx-\xx')\delta(t-t').
\end{equation}
We use the Itô convention for the multiplicative noise in equation~(\ref{eq:particle_current}) and throughout this article~\cite{Dean1996,Oksendal2000}.

The force on the particles of the species $\alpha$ is the sum of the force exerted by the external field and the force due to pair interactions
\begin{equation}
	\ff_\alpha=z_\alpha q \boldsymbol{E} - \sum_\beta \nabla V_{\alpha\beta}*\rho_\beta,
\end{equation}
where $V_{\alpha\beta}(\xx)=q^2 z_\alpha z_\beta/(4\pi\varepsilon r)$ is the electrostatic interaction, with $r=|\xx|$, $\varepsilon$ the dielectric permittivity of the solvent, and $*$ the convolution operator.

We assume that the fluid velocity field $\uu(\xx,t)$ satisfies the fluctuating Stokes equation for incompressible fluids~\cite{DeZarate2006Hydrodynamic} (section~3.2)
\begin{eqnarray}
	\nabla\cdot \boldsymbol{u}  = 0 \label{eq:incompress}\\
	\eta\nabla^2 \boldsymbol{u} - \nabla p = -\sum_\alpha \rho_\alpha \ff_\alpha-\sqrt{\eta T} \nabla\cdot \left(\nnu +\nnu\transp\right), \label{eq:stokes}
\end{eqnarray}
where $\nnu(\xx,t)$ is a Gaussian noise tensor field with correlation function
\begin{equation}
	\langle \nu_{ij}(\xx,t) \nu_{kl}(\xx',t') \rangle=\delta_{ik}\delta_{jl}\delta(\xx-\xx')\delta(t-t').
\end{equation}

In particular, we are interested in the evolution of the density-density correlation function $\langle \rho_{\alpha}(\xx,t) \rho_{\beta}(\xx',t') \rangle$ when the electric field is suddenly switched on ($E(t)=E_0 H(t)$, where $H(t)$ is the Heaviside function), or off ($E(t)=E_0 H(-t)$).
In the first case, the system goes from equilibrium with $E=0$ to NESS with $E=E_0$; in the second case, it relaxes from NESS to equilibrium.

Moreover, we analyze the total charge current $\boldsymbol{J}= q\sum_{\alpha=1}^{M} z_\alpha \langle \boldsymbol{j}_\alpha \rangle$. Using the electroneutrality of the system and after discarding terms of third order in the fluctuations, one finds~\cite{bonneau2023temporal}
\begin{equation}
	\fl	\JJ = \sigma_0 \EE 
	\,- \,  \sum_{\alpha,\beta} q z_\alpha\kappa_\alpha \int \dd\xx \,\boldsymbol{\nabla} V_{\alpha\beta}(\xx)C_{\alpha\beta}(\xx)
	+ \sum_{\alpha,\beta} q^2 z_\alpha z_\beta \EE \int \dd\xx \, \mcO(\xx)C_{\alpha\beta}(\xx).
	\label{eq:current_correlations}
\end{equation}

The correction to the bare current, $\sigma_0\EE$, with $\sigma_0= q^2\sum_\alpha z_\alpha^2 \kappa_\alpha \bar\rho_\alpha$ is the sum of two contributions. The first term involves the electrostatic potential $V_{\alpha\beta}$,
and is referred to 
as the \emph{electrostatic correction}. The second term is called the \emph{hydrodynamic correction} and it involves the Oseen tensor $\mcO$ and the external field $\EE$. 
It.

For completeness we repeat here the approach developed in Ref.~\cite{Demery2016Conductivity,bonneau2023temporal}. The SDFT model presented can be reduced, under the linearization of the field equations, to the following equation for the density fluctuations $n_\alpha={\rho}_{\alpha}-\bar{\rho}_{\alpha}$. In Fourier space, the dynamics of $n_\alpha$ takes the form
\begin{equation}\label{eq:dyn_lin_fourier}
\fl	\dot{\tilde n}_{\alpha}= -\kappa_{\alpha}T k^2 \tilde n_{\alpha} - i \kappa_{\alpha}qz_\alpha \boldsymbol{E}\cdot \boldsymbol{k} \tilde n_\alpha   
	-\kappa_{\alpha}\bar\rho_\alpha k^2 \sum_\beta \tilde V_{\alpha\beta} \tilde n_\beta    +
	\sqrt{\kappa_{\alpha}T\bar{\rho}_{\alpha}} i\kk\cdot\tilde{\zzeta}_{\alpha}.
\end{equation}

Equation~(\ref{eq:dyn_lin_fourier}) allows to formulate an equation for $\langle \tilde n_\alpha(\kk,t) \tilde n_\beta(\kk',t)\rangle= (2\pi)^d \delta(\kk+\kk')\tilde C_{\alpha\beta} (\kk,t)$, the density-density correlation functions
\begin{equation}
	\dot{\tilde C}
	= 2TR -RA\tilde C- \tilde C A^* R,
	\label{eq:correlation_1}
\end{equation}
where $R_{\alpha\beta}(\kk)=\delta_{\alpha\beta}  \bar\rho_\alpha \kappa_\alpha k^2$ and
\begin{equation}
	A_{\alpha\beta}(\kk) = \delta_{\alpha\beta} \frac{T}{\bar\rho_\alpha}\left(1+i\frac{z_\alpha q \boldsymbol{E}\cdot \boldsymbol{k}}{Tk^2}\right) + \tilde V_{\alpha\beta}.
\end{equation}

We can solve equation~(\ref{eq:correlation_1}) exactly for an external field quench since it is a set of linear ODEs with constant coefficients in the interval $[0,t]$.

To further explore the general formulas presented here, 
we focus on the specific case of binary symmetric electrolytes. An ionic solution classified as such is one in which there are only two species of particles with opposite charges and the same mobility. This simplification is of interest for two main reasons. First, many ionic solutions in practical use are very close to binary symmetric solutions \cite{bockris2006}. Secondly, this simplification allows us to manipulate analytically and derive simple form expressions that still capture the system's global behavior.

 A binary symmetric electrolyte, where both species have the same mobility and charge satisfy $\alpha=\{+,-\}$, $\, z_+=-z_-=1$, $\,\bar\rho_\alpha=\bar\rho$, and $\kappa_\alpha=\kappa$.
We nondimensionalize equation~(\ref{eq:correlation_1}) by setting $\tilde{C}= \bar \rho \tilde c$ and $\kk=\ss/\lambda\ind{D}$ where $\ldeb=\sqrt{T \varepsilon/(2q^2 \bar{\rho})}$ is the Debye length. 
Then we rescale time by the Debye time $t\ind{D} = \ldeb^2 /(\kappa T)$, $t=t\ind{D}\tau$. 
We rewrite the external field to separate the magnitude from the time dependence $E(t) = E_0 g(t)$ and introduce the dimensionless field $f= q \ldeb E_0/T$.
The rescaled correlation $c_{\alpha\beta}(\tau)$ follows
\begin{equation}\label{eq:evol_cor_adim}
	\dot {\tilde c} = 2s^2-\omega \tilde c-\tilde c\omega^*,
\end{equation}
where we have introduced the matrix $\omega_{\alpha\beta}(\ss) = \delta_{\alpha\beta}\left(s^2+iz_\alpha fs_\parallel \right)+\frac{z_\alpha z_\beta}{2}$.

\section{Correlations in electrolyte systems -- Long range behavior at NESS}
\label{sec:stationary_case}

In this section, we analyze the structure of the correlations between different ions at scales larger than the Debye length. 
The correlation elements in Fourier space for binary symmetric electrolytes can be calculated using equation~(\ref{eq:evol_cor_adim}). In the stationary state case, as shown in \cite{Demery2016Conductivity}, the equation reduces to a set of algebraic equations to which the solution is
\begin{eqnarray}\label{eq:NESS_corr}
	\fl \tilde{c}(\ss) =\frac{s^2}{2 \left(1+2 s^2\right) \left(s^2+s^4+f^2 s_\parallel^2\right)} \nonumber \\
	\times \left(
	\begin{array}{cc}
		1+4 s^2 +4 s^4+4 f^2 s_\parallel^2&1+2 s^2 -2 i f s_\parallel \\
		1+2 s^2 +2 i f s_\parallel& 1+4 s^2 +4 s^4+4 f^2 s_\parallel^2 \\
	\end{array}
	\right).
\end{eqnarray}

\begin{figure}
	\centering
	\includegraphics[width=0.9\linewidth]{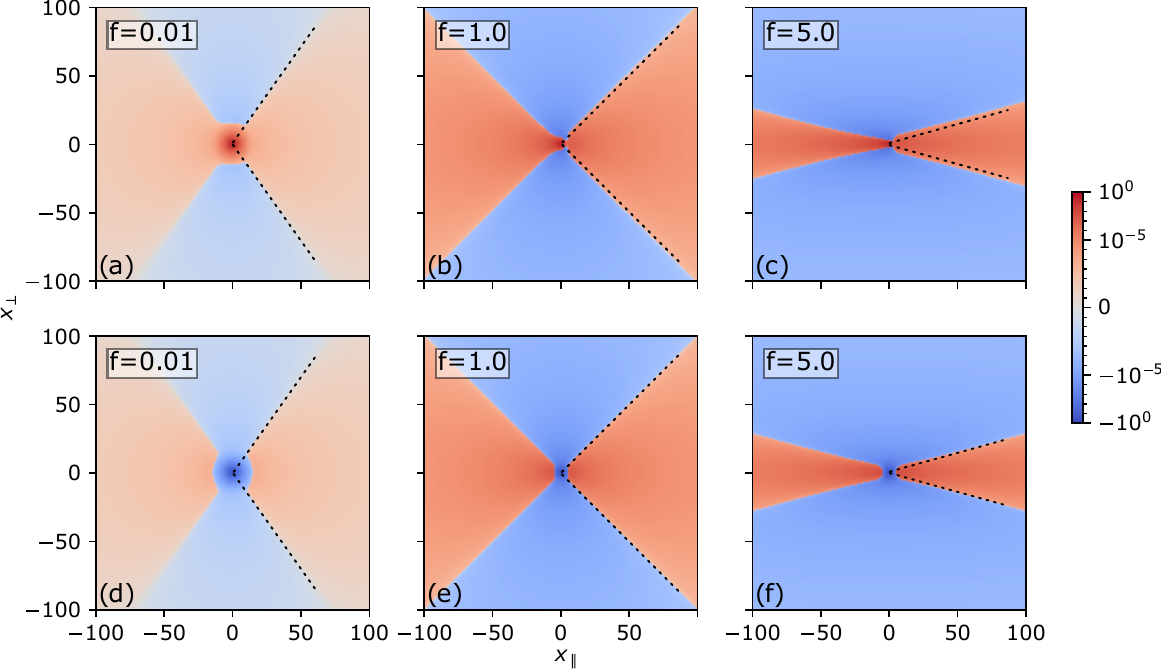}
	\caption{The correlation function in the non-equilibrium stationary state for different values of the normalized external field $f$. Panels (a,b,c) are the anion-cation correlation $c_{-+}$, and panels (d,e,f) are the equal charge correlation $c_{\alpha\alpha}$.    One can observe that far from the origin, a cone develops and that at vanishing fields, the angle of the cone reaches a finite value, as predicted in equation~(\ref{eq:cone_angle}). Surprisingly, the long-range shape of any ion-ion correlation function is the same. In dashed black line is the prediction of equation~(\ref{eq:cone_angle}).}	
	\label{fig:NESS_3_values}
\end{figure}

In figure~\ref{fig:NESS_3_values} one can see the NESS correlation elements in real space for different values of the dimensionless field, obtained from the numerical Fourier inversion of equation~(\ref{eq:NESS_corr}).
A clear conical structure with respect to the field direction is visible for both $c_{++}$ and $c_{-+}$. 
The head angle of the cone decreases with the electric field, but does not reach $\pi/2$ as the field goes to zero.
Differences between $c_{++}$ and $c_{-+}$ are manifested only around the origin at distances comparable with $\ldeb$.

\subsection{Discontinuity at small wave number}
Examining $\tilde c(\ss)$, we see that there is a discontinuity at the origin, since the limit is different upon approaching the origin from the direction perpendicular or parallel to the electric field
\begin{equation}
	\fl \tilde{c}_{\alpha\beta}\left(s_\parallel=0,s_\perp\right) \underset{s_{\perp} \to 0}{ = } \frac{1}{2}\,,  \ \quad 	\tilde{c}_{\alpha\beta}\left(s_{\parallel},s_{\perp}=0\right) \underset{s_\parallel \to 0}{=} \frac{1}{2\left(1+f^2\right)}\,.
\end{equation}
The fact that the Fourier transform of the correlation function is not continuous at the origin usually indicates an algebraic decay in real space. To determine the long-range properties of the correlation function, we can separate this discontinuity from the full expression. 
By doing so, we are left with a term called the singular part $\tilde{c}\ind{sing}$, which encompasses the discontinuity and accounts for the long-range behavior in real space, and the rest, namely $\tilde{c} - \tilde{c}\ind{sing}$, which is more regular at the origin and hence, its contribution decays faster in real space.
Note that the singular part is defined up to a regular function. 
We can extract the discontinuity from equation~(\ref{eq:NESS_corr}) by discarding higher powers of $s_{\perp}$ and $s_\parallel$ for the numerator and the denominator separately.
In this way, we find the singular part, which is identical for all correlation elements
\begin{equation}
	\label{eq:c_singular_fourier}
	\tilde{c}\ind{sing} = \frac{s^2}{2\left(s^2+f^2s_\parallel^2\right)}.
\end{equation}
One can verify that subtracting this term from the correlation elements regularizes the behavior at the origin, and the regularized term goes to 0 in the $s\to0$ limit.
Therefore, the inverse Fourier transform of the singular part dominates the behavior of the correlation elements at long distances in real space. We can invert the Fourier transform of the singular part
and express it in terms of the
Green function of the Poisson equation $G$
\begin{equation}
	c\ind{sing} = - \frac{1}{2\left(2\pi\right)^d\sqrt{1+f^2}} \nabla^2_x G\left(\hat{x}\right).
\end{equation}
where $\hat{\xx}=\left\{x_\parallel /\sqrt{1+f^2},\xx_\perp\right\}$.
This expression can be evaluated in $d$ dimensions to get
\begin{eqnarray}\label{eq:corr_long_range_ness_d}
\fl	c\ind{sing} =	-\frac{f^2}{4\pi ^{d/2}}   \left(f^2+1\right)^{\frac{d-3}{2}} \Gamma \left(\frac{d}{2}\right)g_d
	\left(\sqrt{1+f^2}\frac{x_\perp}{x_\parallel}\right) \frac{1}{x_\parallel^d	}\nonumber\\  \mathrm{with} \ \ g_d(y) =  \frac{y^2-(d-1) }{\left(y^2+1\right)^{\frac{d}{2}+1} },
\end{eqnarray}
where $\Gamma$ is the gamma function. 
For $d=3$, it reduces to
\begin{equation}\label{eq:corr_scaling_real}
	c\ind{sing} = -\frac{f^2}{8 \pi  x_\parallel^3}\ g_3\left(\sqrt{1+f^2}\frac{x_\perp}{x_\parallel}\right)\ \ \mathrm{with} \ \ g_3(y) =  \frac{y^2-2}{ \left(y^2+1\right)^{5/2}}.
\end{equation}
The expression in equation~(\ref{eq:corr_long_range_ness_d}) has several interesting features. First, the argument of the function $g_d$ contains the ratio between the $x_\parallel$ and $x_\perp$ coordinates. This implies a conical shape around the $x_\parallel$ axis, which is observed in figure~\ref{fig:NESS_3_values}. This is an intrinsic difference from systems with short-range interactions, where the correlations have a parabolic shape \cite{poncet2017,bain2017critical}. 
Second, the angle of the cone does not approach $\frac{\pi}{2}$, as one might expect, when the field is decreased towards zero. Instead, it settles at an angle, unlike, for example, the Mach cone.
From equation~(\ref{eq:corr_long_range_ness_d}), we can see that the angle $\Theta$ of the cone where the value of the correlation function is 0 is given by
\begin{equation}\label{eq:cone_angle}
	\Theta_d = \sin^{-1}\left(\sqrt{\frac{d-1}{d+f^2}}\right).
\end{equation}
The expression in equation~(\ref{eq:cone_angle}) shows that the angle of the cone in any dimensions 
converges to $\sin^{-1}\left(\sqrt{1-1/d}\right)$ as the field goes to $0$ (figure \ref{fig:NESS_3_values}). 
In other words, at large distances, the first non-zero term of the correlation in powers of $f$ is not spherically symmetric. 
\begin{figure}
	\centering
	\includegraphics[width=0.8\linewidth]{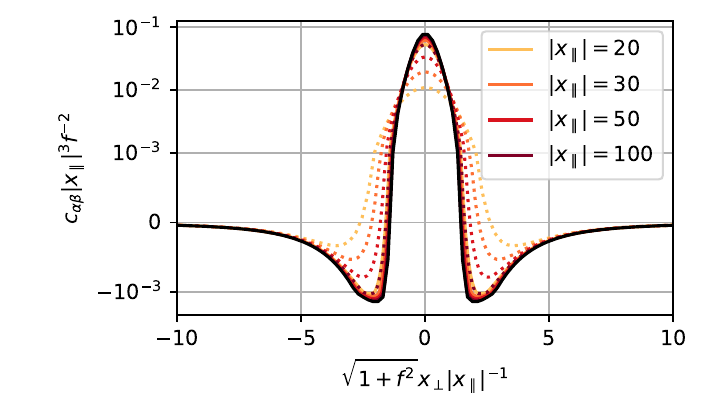}
	\caption{The rescaled correlation function  $C_{\alpha\beta}|x_\parallel|^3f^{-2}$ along slices of constant $x_\parallel$ values. This allows us to compare it to the scaling function $g_3$. The dotted lines correspond to $f=1$ while the solid lines correspond to $f=5$. We can see in solid black the evaluation of equation~(\ref{eq:corr_scaling_real}).}	
	\label{fig:scaled_corr}
\end{figure}
Finally, the structure of the correlation in equation~(\ref{eq:corr_long_range_ness_d}) shows a self-similar shape for each cut along the $x_\parallel$ axis. In fact, by stretching this scaling function $g_d$ by the factor $\sqrt{1+f^2}$, we find a universal shape for any cut and external field.
The function $g_3$ is presented together with a numerical inversion of the Fourier transform of the correlation element in figure~\ref{fig:scaled_corr}. 

We remark that the correlation shape at large scales is similar in structure to the electric potential generated by a simple quadrupole moment charge configuration (equally spaced collinear charges of $+q,-2q,+q$ along $x_\parallel$) where $f^2$ is analogous to the magnitude of the quadrupole moment tensor. This similarity suggests a relation to the picture of the ionic cloud deformation proposed in the works of Onsager \cite{Onsager1957Wien,onsager1939}.

\section{Time dependent correlations}\label{sec:time dependent}

Now that we have characterized the correlations in the NESS, we investigate the temporal evolution of the correlations between the short-ranged correlations at equilibrium and the algebraic correlations in the NESS.
The focus remains on lengths that are large compared to the Debye length and times that are large compared to the Debye time.

\subsection{Switching on of the external field}

Equation~(\ref{eq:evol_cor_adim}) can be solved exactly for an immediate switching on of the electric field, namely, $E(t)=E_0 H(t)$ to give
\begin{eqnarray}
\fl	\tilde c_{++}  =
	\frac{f^2 u^2 e^{-B \tau}}{2 A B C \Delta }
	\left[B^2 \cosh \left(\sqrt{\Delta } \tau\right)+B \sqrt{\Delta } \sinh \left(\sqrt{\Delta } \tau\right)-4 C s^2\right] + \frac{1}{2 B C}-\frac{1}{B}+1,\nonumber\\ \label{eq:corr_exact_++}\\
\fl	\tilde c_{+-} =\frac{f u e^{-B \tau}}{2 A B C \Delta } \Big[B \sqrt{\Delta } f u \sinh \left(\sqrt{\Delta } \tau\right) (B-2 i f s u)+B f u \cosh \left(\sqrt{\Delta } \tau\right) (\Delta -2 i B f s u) \nonumber\\
+2 i C s\Big]+\frac{1}{2 C}-\frac{i f s u}{B C},\label{eq:corr_exact_+-}
\end{eqnarray}
we have introduced the variables $u=s_\parallel/s$, $A=1+s^2$, $B=1+2s^2$, $C=f^2 u^2+s^2+1$ and $\Delta= 1-4f^2s^2u^2$.
The other terms are deduced with $\tilde c_{--}=\tilde c_{++}$ and $\tilde c_{-+}=\tilde c_{+-}^*$.
At long times, the dominant, time-dependent term in equations~(\ref{eq:corr_exact_++}) and (\ref{eq:corr_exact_+-}) corresponds to the smallest eigenvalue of equation~(\ref{eq:evol_cor_adim}), $B-\sqrt{\Delta}$.
We can read it in the expressions in equations~(\ref{eq:corr_exact_++}) and (\ref{eq:corr_exact_+-})
\begin{eqnarray}
\fl	\tilde c_{++} \underset{\tau\to\infty}{\sim}
	\frac{f^2 u^2 e^{- \tau(B-\sqrt{\Delta})}}{4 A  C \Delta }
	\left[B + \sqrt{\Delta } \right] + \frac{1}{2 B C}-\frac{1}{B}+1,\\
\fl	\tilde c_{+-} \underset{\tau\to\infty}{\sim}\frac{f^2 u^2 e^{- \tau(B-\sqrt{\Delta})}}{4 A C \Delta } \left[ \sqrt{\Delta }   (B-2 i f s u)+   (\Delta -2 i B f s u)\right]
	+\frac{1}{2 C}-\frac{i f s u}{B C}.
\end{eqnarray}
These correlation terms can be approximated in the long-range regime, which translates to $s\to0$ in non-dimensionalized Fourier space. Again, we find that the behavior of all the correlation terms is the same
\begin{equation}\label{eq:corr_long_range_fourier}
	\tilde c_{\alpha\beta} \sim
	\frac{f^2 s_\parallel^2 }{2   \left(s^2+f^2s_\parallel^2\right)  }e^{- 2\tau\left(s^2+f^2s_\parallel^2\right)}
	+ \frac{s^2}{2 \left(s^2+f^2s_\parallel^2\right) },
\end{equation}
One sees that when $\tau\to0$, the correlation is $1/2$. This is the Yukawa correlation found in equilibrium: Seen at large distances, it is a delta function at the origin, which gives a constant in Fourier space. When $\tau\to\infty$ one recovers the NESS result from the previous section. 
We can identify the exponential as the solution to the diffusion equation in Fourier space, which means that the transition between these two states follows diffusive dynamics. The term $f^2s_\parallel^2$ enhances the diffusion in the direction of the driving field. 
The correlation spreads like a nonisotropic diffusion process with diffusion constants $\kappa T$ in the perpendicular directions and $ \kappa T(1+f^2)$ in the field direction. 

We can invert the time-dependent term back to real space to get
\begin{equation}\label{eq:corr_long_range_d}
	c(x,\tau) =-\frac{f^2}{8\pi^{d/2}  \tau^{d/2} \left(1+f^2\right)^{3/2}} \frac{\partial^2}{\partial \bar{x}_\parallel^2} \left[ \frac{\Gamma_{d/2-1}\left(\frac{\bar{x}^2}{8}\right)}{\bar{x}^{d-2}} 	\right],
\end{equation}
where $\Gamma_{d/2-1}$ is the (upper) incomplete gamma function of $d/2-1$ and $\bar{x}=\left\{\frac{x_\parallel}{\sqrt{\tau}\sqrt{1+f^2}},\frac{\xx_\perp}{\sqrt{\tau}}\right\}$. This shape of the correlation can be written as
\begin{equation}\label{eq:c_of_phi}
	c(x,\tau) = \frac{1}{8 \pi^{d/2}}\frac{1}{\tau^{d/2}} \frac{f^2}{(1+f^2)^{3/2}}\Phi\left(\frac{\xx}{\sqrt{\tau}}\right).
\end{equation}
The function $\Phi$ is anisotropic in space and its asymptotic behavior is given by
\begin{eqnarray}
\fl	\Phi(\boldsymbol{y})=2 \left(f^2+1\right)^{d/2}  \Gamma \left(\frac{d}{2}\right)\frac{\left(d-1\right)y^2_\parallel-\left(1+f^2\right)y^2_\perp}{\left(y^2_\parallel+\left(1+f^2\right)y^2_\perp\right)^{\frac{d}{2}+1}} \sim \frac{1}{y^d}\, \ \mathrm{ when } \, \ y\to0, \\
\fl	\Phi(\boldsymbol{y})= \frac{y^2_\parallel}{y^2_\parallel+\left(1+f^2\right)y^2_\perp}\exp\left(-\frac{y^2_\parallel+\left(1+f^2\right)y^2_\perp}{8 \left(1+f^2\right)}\right)\sim \exp\left(-y^2 \right)\, \ \mathrm{ when }\, \  y\to\infty.
\end{eqnarray}

\begin{figure} 
	\centering
	\includegraphics[width=0.6\linewidth]{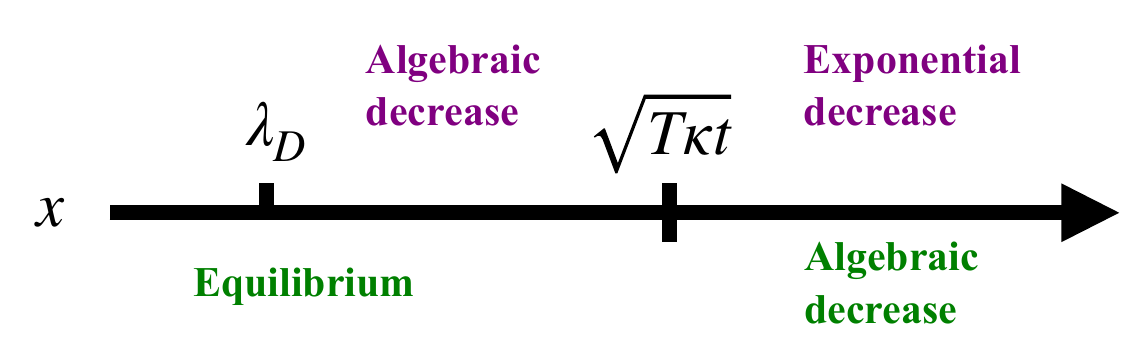}
	\caption{Sketch of the length scales at play and the functional dependence of the density-density correlation function on the distance from the origin. The behavior in the upper row (magenta) corresponds to the spatial relaxation of the correlation function from equilibrium to NESS (switching on). The behavior in the lower row (green) corresponds to the spatial relaxation of the correlation function from NESS to equilibrium (switching off).}	
	\label{fig:length_scales}
\end{figure}

\begin{figure}
	\centering
	\includegraphics[width=1\linewidth]{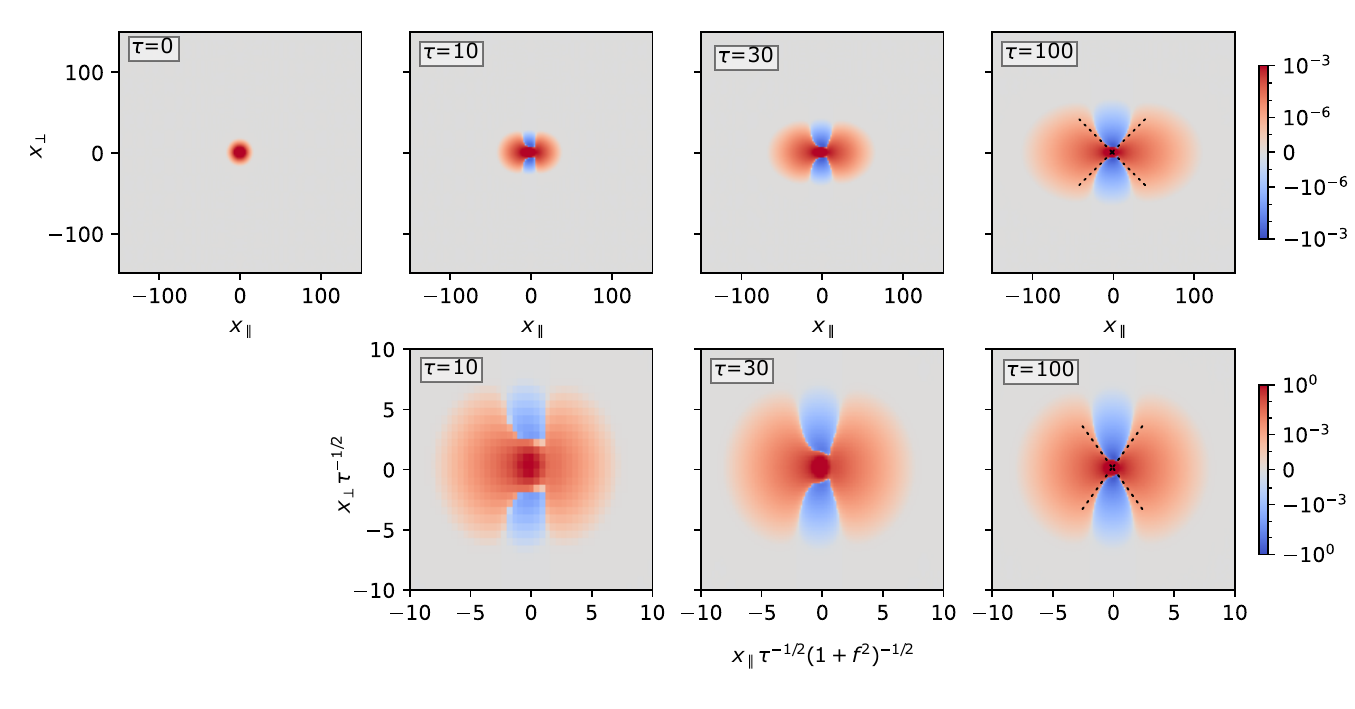}
	\caption{
		A time series of the correlation element  $c_{-+} $ after a sudden switching on of the external field.
		In the upper row, $c_{-+} $ along real axes. In the lower row,  
		the rescaled correlation function $c_{-+} \tau^{3/2}$ along rescaled axes. 
		The figure has been evaluated for $f=1$. Near the center, one can see the conical shape found in the NESS. Away from the center, we find the exponentially small equilibrium value. 
		In the rightmost panels, the angle predicted by equation~(\ref{eq:cone_angle}) was added, allowing us to appreciate the different behavior regimes visually. }  	
	\label{fig:corr_time_series}
\end{figure}

The argument of $\Phi$ in equation~(\ref{eq:c_of_phi}) presents diffusive scaling between the spatial and temporal coordinates.
This indicates that the observation length should be compared to the length $\sqrt{\tau}$. 
At a given time, the NESS correlations are observed below the length $\sqrt{T\kappa t}$ while an exponential decay of the correlations in space is observed beyond it (figure~\ref{fig:length_scales}).
In figure~\ref{fig:corr_time_series}, one can see the different regimes in rescaled and absolute axes at different times. 
\subsection{Switching off of the external field}
A similar procedure can be applied for the switch-off of the field. Equation~(\ref{eq:evol_cor_adim}) can be solved exactly for an immediate switching off of the electric field, to give
\begin{eqnarray}
	\tilde c_{++} =
	-\frac{1}{2 A}	-\frac{f^2 u^2 e^{-B \tau}}{2 A B C} [B \sinh (\tau)+\cosh (\tau)],\\
	\tilde c_{+-} =-\frac{f^2 s^2 u^2 e^{-2 A \tau}}{2 A B C}-\frac{f^2 u^2 e^{-2 s^2 \tau}}{2 B C}-\frac{i f s u e^{-B \tau}}{B C}+\frac{1}{2 A},
\end{eqnarray}
At large times, the dominant terms of each correlation element are 
\begin{eqnarray}
	\tilde c_{++} =
	-\frac{1}{2 \left(1+s^2\right)}	-\frac{f^2 u^2  e^{-2 s^2 \tau}}{2  \left(1+2s^2\right) \left(1+s^2+f^2 u^2\right)},\\
	\tilde c_{+-} =\frac{1}{2 \left(1+s^2\right)}	-\frac{f^2 u^2  e^{-2 s^2 \tau}}{2  \left(1+2s^2\right) \left(1+s^2+f^2 u^2\right)}.
\end{eqnarray}

\begin{figure}
	\centering
	\includegraphics[width=1\linewidth]{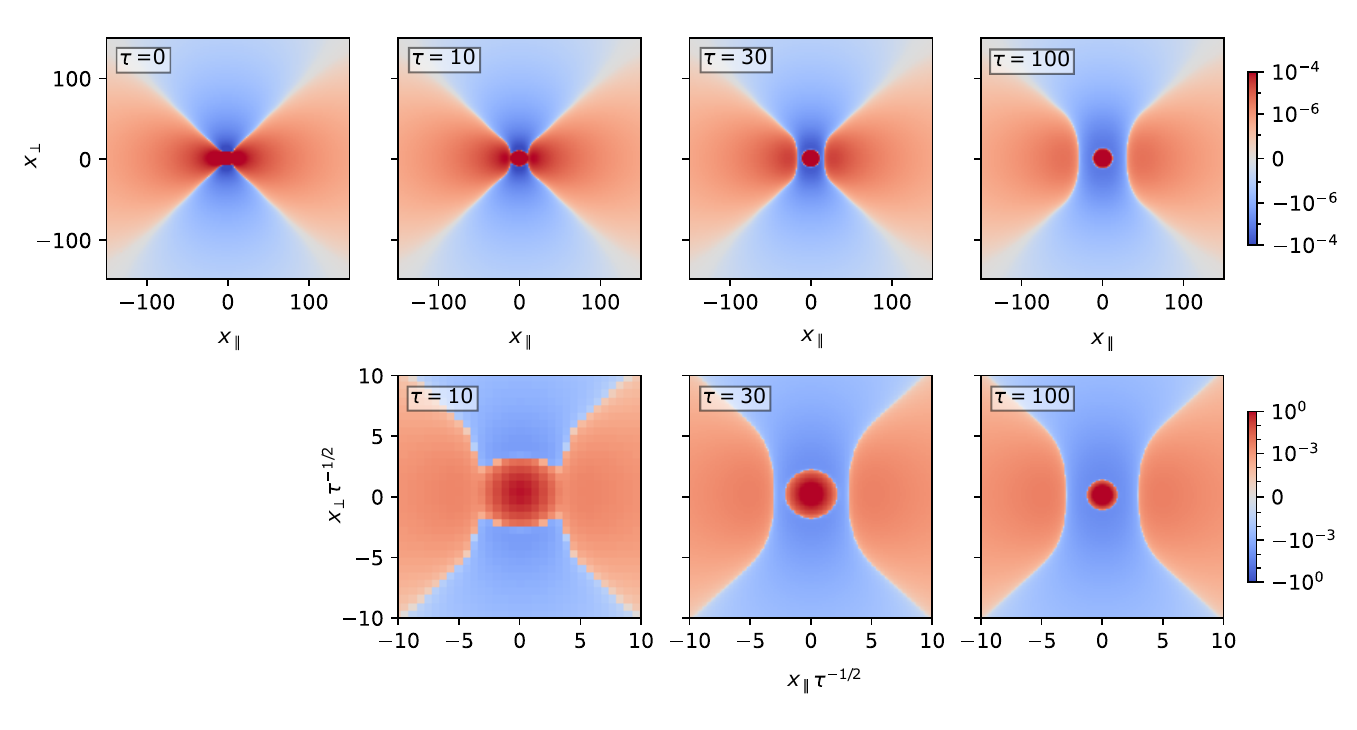}
	\caption{A time series of the correlation element $c_{-+} $ after a sudden switching off of the external field.
	In the upper row, $c_{-+} $ along real axes. In the lower row,  
	the rescaled correlation function $c_{-+} \tau^{3/2}$ along rescaled axes. 
	The figure has been evaluated for $f=1$. Away from the center, one can see the conical shape found in the NESS. Near the center, we find the exponentially small equilibrium values. }  	
	\label{fig:switch_off}
\end{figure}

These correlation terms can be approximated in the long-range regime, which translates to $s\to0$ in non-dimensionalized Fourier space. Again, we find that the behavior of all the correlation terms is the same
\begin{eqnarray}
	\tilde c_{++} \underset{s\to0}{\sim}
	-\frac{1}{2 }	-\frac{f^2 s_\parallel^2  e^{-2 s^2 \tau}}{2   \left(s^2+f^2 s_\parallel^2\right)},\\
	\tilde c_{+-} \underset{s\to0}{\sim} \frac{1}{2 }	-\frac{f^2 s_\parallel^2  e^{-2 s^2 \tau}}{2   \left(s^2+f^2 s_\parallel^2\right)},
\end{eqnarray}
One sees that when $\tau\to0$, the correlation is the NESS solution at large distances (equation~(\ref{eq:c_singular_fourier})). When $\tau\to\infty$, the equilibrium value of $1/2$ is recovered (See figure~\ref{fig:switch_off}).
In real space, the dynamics are given by convolution between the NESS solution at a large scale and the diffusion equation fundamental solution, which can be written as
\begin{equation}
	c(x,\tau) = \frac{f^2}{2\left(2\pi \right)^d \tau^{d/2} } \Psi\left(\frac{\xx}{\sqrt{\tau}}\right),
\end{equation}
with $ \Psi(\xx)=\int \frac{e^{-2\hat{s}^2}}{\hat{s}^2+f^2\hat{s}_\parallel^2}\hat{s}^2_\parallel e^{-i \hat{\ss}\xx} \dd\hat{\ss}$. The diffusive nature, manifested in the scaling of the dynamics, is also preserved in this case.

Figure~\ref{fig:switch_off} presents the relaxation of $c_{-+}$ after a switching off of the external field. 
At a given time, the equilibrium correlation function, of the spherically symmetric Yukawa form, is observed below the length $\sqrt{T\kappa t}$ while the NESS conical correlation with an algebraic decay in space is observed beyond it. 

\section{Mesoscopic density fields and conductivity}\label{sec:eq for long range}
The diffusive dynamics of the correlation suggest a relation to the algebraic relaxation of the charge current reported in~\cite{bonneau2023temporal}. To establish this relation, we start by normalizing the expression in equation~(\ref{eq:current_correlations}) to get the dimensionless total charge current
\begin{equation}
	\frac{J(\tau)}{\sigma_0 E_0} =g(\tau)+\frac{ 1 }{\bar{\rho} \ldeb^3  } \gamma\ind{el}(\tau,f)+g(\tau)\frac{r_s}{\ldeb} \gamma\ind{hyd}(\tau,f),
	\label{eq:corrections}
\end{equation}
where $r_s=(6\pi\eta\kappa)^{-1}$ is the hydrodynamic radius of the charged particles and $g(\tau)=E(\tau)/E_0$ is the temporal dependence of the external electric field. 
The electrostatic and hydrodynamic corrections $\gamma\ind{el}$ and $\gamma\ind{hyd}$ in equation (\ref{eq:corrections}) read, respectively
\begin{eqnarray}
	\gamma\ind{el}(\tau,f) = - \frac{1}{32 \pi^3 f} \int  \dd \ss  \frac{s_\parallel}{s^2}i \sum_{\alpha,\beta}z_\alpha \left[\tilde c_{\alpha\beta}(\ss,\tau,f)-\delta_{\alpha\beta}\right], \label{eq:gamma_el}\\
	\gamma\ind{hyd}(\tau,f) = \frac{3}{8 \pi^2 } \int  \dd \ss \frac{1}{s^2} \left( 1- \frac{s^2_\parallel}{s^2} \right)\sum_{\alpha,\beta}z_\alpha z_\beta \left[\tilde c_{\alpha\beta}(\ss,\tau,f)-\delta_{\alpha\beta}\right]. \label{eq:gamma_hyd}
\end{eqnarray}

Examining the integrals in equations~(\ref{eq:gamma_el}) and (\ref{eq:gamma_hyd}),
we can see that the results for the correlation elements found in section~\ref{sec:time dependent}  are not sufficient to describe the dynamics of the charge current corrections: All the correlation elements are equal and cancel in the sums. Hence, some finer details of the correlation elements are needed.

For a binary symmetric electrolyte, the summation terms in the integrals in equations~(\ref{eq:gamma_el}) and (\ref{eq:gamma_hyd})
suggest defining new field variables, the number density $U=\tilde n_++\tilde n_-$ and the charge $\Delta=\tilde n_+-\tilde n_-$. With these variables the correction terms can be written as
\begin{eqnarray}
	\gamma\ind{el}(\tau,f) = - \frac{1}{32 \pi^3 f} \int  \dd \ss  \frac{s_\parallel}{s^2} i\, \tilde{c}_{U\Delta}(\ss,\tau,f), \\ 
	\gamma\ind{hyd}(\tau,f) = \frac{3}{8 \pi^2 } \int  \dd \ss \frac{1}{s^2} \left( 1- \frac{s^2_\parallel}{s^2} \right)\tilde{c}_{\Delta\Delta}(\ss,\tau,f). \label{eq:gamma_hyd2}
\end{eqnarray}

In the following, we examine the dynamics and correlation of $U$ and $\Delta$ to understand the temporal relaxation of the total charge current.

\subsection{Mesoscopic equations}

Similarly to \cite{Mahdisoltani2021Transient}, we start by considering equation~(\ref{eq:dyn_lin_fourier}) for the density of the species $\alpha$ in Fourier space. For the particular case of binary symmetric electrolytes, 
we can write the equations for the number density and charge fields
\begin{eqnarray}
	\dot{U}&= -\kappa T k^2 U - i \kappa q  \boldsymbol{E}\cdot \boldsymbol{k} \Delta   
	+\sqrt{2} \chi_{U},\label{eq:U} \\
	\dot{\Delta}&= -\kappa T k^2  \Delta - i \kappa q  \boldsymbol{E}\cdot \boldsymbol{k} U   
	-2 \kappa \bar\rho\frac{q^2}{\varepsilon}  \Delta   
	+\sqrt{2} \chi_{\Delta}.	\label{eq:delta}
\end{eqnarray}

When $E$ is set to $0$, the two equations decouple, equation~(\ref{eq:U}) describes (noisy) diffusion, and equation~(\ref{eq:delta}) describes a (noisy) diffusion in the presence of screening, which flattens even the slow modes of the fluctuations after a Debye time.

Applying a field couples the equations; in particular, this coupling gives rise to a charge fluctuation that persists beyond the Debye time.
To address the asymptotic long-distance behavior of a system, it is often helpful to simplify the analysis by identifying specific regimes where particular physical processes dominate the dynamics. In the context of charge fluctuations, for example, it is known that the behavior is strongly influenced by the Debye screening mechanism at short length and time scales. However, when longer lengths and time scales are considered, it is often possible to approximate the charge fluctuations with a quasi-stationary solution that captures the dominant features of the system. 

Following this logic, we examine the behavior of equations~(\ref{eq:U}) and (\ref{eq:delta}) 
at long times 
compared to $t\ind{D}$, which allows us to neglect the temporal derivative and at large lengths compared to $\lambda\ind{D}$, which allows to neglect the diffusive and noise terms. In this regime, equation~(\ref{eq:delta}) simplifies to
\begin{equation}\label{eq:delta_aprox}
	2 \kappa \bar\rho\frac{q^2}{\varepsilon}  \Delta =  -i \kappa q  \boldsymbol{E}\cdot \boldsymbol{k} U.
\end{equation}

Note that the noise term is also of higher order in $\lambda_D$.
Now we can use equations~(\ref{eq:U}) and (\ref{eq:delta_aprox}) to write a closed equation for $U$
\begin{equation}\label{eq:U_2}
	\dot{U}= -\kappa T \left(k^2+f^2 k_\parallel^2\right) U    
	+\sqrt{2} \chi_{U}.
\end{equation}

We can write an equivalent version of equation~(\ref{eq:correlation_1}) that is valid only for large distances compared to the Debye length. Similarly, we have $\langle U(k,t)U(k',t)\rangle=(2\pi)^d\delta(k+k')C_{UU}(k,t)$, where $C_{UU}(k,t)$ is the Fourier transform of the correlation function in real space. $C_{UU}$ satisfies a diffusion equation with a source term
\begin{equation}
	\dot{C }_{UU}(k,t)= -2\kappa T \left(k^2+f^2 k_\parallel^2\right) C_{UU}(k,t)     
	+4  \kappa T  \bar{\rho} k^2. 
\end{equation}
Nondimensionlizing the equation in the same way done in section~\ref{sec:model}
 gives
\begin{equation}\label{eq:c_U}
	\dot{c }_{UU}(s,\tau)= - 2 \left(s^2+f^2 s_\parallel^2\right) c_{UU}(s,\tau)     
	+4 s^2 ,
\end{equation}
we recall that $c_{UU} = \sum_{\alpha\beta} \tilde{c}_{\alpha\beta}$ 
with the initial condition $C_{sp}(s,0) = \frac{1}{2}$. It is easy to see that equation~(\ref{eq:corr_long_range_fourier}) gives the solution to this equation.
At this stage, we can gain some insight into the relaxation rates of the conductivity corrections that we found in \cite{bonneau2023temporal}. 

\subsection{Relaxation rates of the currents}\label{subsec: relaxation}
When we swich off the external field, the system transitions from NESS to equilibrium. Examining equations~(\ref{eq:U}) and (\ref{eq:delta}) we see that when the field $E$ is set to 0, the equations decouple and fluctuations in $\Delta$ decay quickly over a time scale $t\ind{D}$. This leads to a rapid decay of the correlation function between $U$ and $\Delta$  and the autocorrelation of $\Delta$.
This is the origin of the exponential decay of the total electric current upon switching off the external field. 

To understand the second transition, from equilibrium to NESS, we see that with equation~(\ref{eq:delta_aprox}) one can express $c_{U\Delta}$ and $c_{\Delta\Delta}$ with the solution to equation~(\ref{eq:c_U}). The time-dependent parts of $c_{U\Delta}$ and $c_{\Delta\Delta}$ are
\begin{eqnarray}
	\delta c_{U\Delta}(s,\tau)= -\frac{2 i  f^3 s_\parallel^3 }{   s^2+f^2s_\parallel^2  }e^{- 2\tau\left(s^2+f^2s_\parallel^2\right)},
		\\	
	\delta c_{\Delta\Delta}(s,\tau)= \frac{2   f^4 s_\parallel^4 }{   s^2+f^2s_\parallel^2  }e^{- 2\tau\left(s^2+f^2s_\parallel^2\right)}.
\end{eqnarray}
Now we can compute the relaxation terms of the currents in equations~(\ref{eq:gamma_el}) and (\ref{eq:gamma_hyd}) 
\begin{eqnarray}
	\delta \gamma\ind{el}(\tau,f) =  - \frac{3 \left(f^2+1\right)^{3/2} \sinh ^{-1}(f)-4 f^3-3 f}{96 \sqrt{2} \pi ^{3/2}   f^3 \left(f^2+1\right)^{3/2} }\frac{1}{\tau^{3/2}}, \\
	\delta \gamma\ind{hyd}(\tau,f) =\frac{1}{16 \sqrt{2 \pi } }\left[ \frac{15+6f^2}{ f^3 }\sinh ^{-1}(f)-\frac{15+11f^2}{ f^2  \sqrt{f^2+1 }}\right]\frac{1}{\tau^{3/2}}, \label{eq:dgamma_hyd2}
\end{eqnarray}
consistent with the results reported  in~\cite{bonneau2023temporal} for the algebraic relaxation rate and prefactor dependence on the external field. 
In other words, the correlation terms respect a diffusive scaling, and due to the fact that the interaction kernels $\tilde{\mcO}\sim \tilde{V}\sim \frac{1}{s^2}$ has a long range algebraic structure, the total charge current relax algebraiclly towards its value in the NESS.

This effect is reminiscent of the {\it long time tails} effect seen in hydrodynamic systems \cite{hansen2013}. The long-time tails phenomenon involves the inertia of the fluid, which is absent in our model.

\section{Conclusion}\label{sec:conclusions}
In this work, we have characterized the behavior of the particle-particle correlation functions in the long-range regime, in the non-equilibrium stationary state (NESS), and in the transient regime as the system approaches NESS.
At NESS, the density-density correlation functions are anisotropic and decay algebraically with distance. These properties persist even in the weak-field limit.
The correlations exhibit a self-similar universal form for cuts along the axis parallel to the external field $x_\parallel$.
This self-similar structure is conical along the $x_\parallel$ axis. This distinguishes ionic systems from systems with short-range interactions. 
In both types of systems (short-range and long-range interactions), particle-particle correlations are short-range at equilibrium and long-range at NESS.
However, the correlation spatial structure is conical and not parabolic.
Moreover, we examined the relaxation of the correlation functions from equilibrium to NESS. It is characterized by a diffusive length scale $\sqrt{T\kappa t}$. At short distances compared with $\sqrt{T\kappa t}$, the NESS correlation function is found. For larger distances, the correlations are exponentially decaying.
Lastly, we approximated the equations for the ionic fluctuation fields to explain the relaxations of the total charge currents towards equilibrium and non-equilibrium stationary state. 

Recently, the temporal correlations of the fluctuations of number and charge densities have been investigated at equilibrium using SDFT~\cite{rotenberg2023ionic}.
It has been found that the number correlations decay with time as $t^{-3/2}$, which was attributed to the diffusion of the ions.
That the same algebraic decay was found for these correlations and the relaxation of the electric current following a sudden switching on of the electric field~\cite{bonneau2023temporal} raises the question of a possible relation between the two phenomena.
However, the algebraic relaxation of the current and the conical correlations in the NESS appear only beyond linear response, so that they cannot be directly connected to equilibrium fluctuations.

\section*{Acknowledgments}

We thank Hélène Berthoumieux, David Andelman and Henri Orland for stimulating discussions.

\vspace{0.5cm}
\providecommand{\newblock}{}

\end{document}